\documentclass[10pt]{article}
\usepackage{pst-all}
\usepackage{letterspace}
\usepackage{psfrag}
\usepackage{latexsym}
\usepackage{amsmath}
\usepackage[dvips]{graphicx}
\usepackage{caption2}

\def\yskip{\penalty-50\vskip3pt plus3pt minus2pt}
\def\y{\yskip}
\def\yy{\yskip\yskip}
\def\yyy{\yskip\yskip\yskip}
\def\qed{\hskip 3pt\vrule height 6pt width 3pt depth 0pt}
\def\q5uad{\quad\quad\quad\quad\quad}

\def\chl{\yyy\y\par\noindent}

\def\s{\ }
\def\ss{\ \ }
\def\sss{\ \ \ }
\def\ssss{\ \ \ \ }
\def\bs{\ssss \s}
\def\Bs{\bs \bs}
\def\BBs{\Bs \Bs}

\usepackage{fancyheadings}
\small

\normalsize

\title{\bf The Number of Spanning Trees in $K_n$-complements of Quasi-threshold Graphs\vspace{0.4cm}}

\author{Stavros D. Nikolopoulos \ and \ Charis Papadopoulos}

\date{}

\begin{document}

\maketitle

\vspace{-0.4cm}

\centerline{\it Department of Computer Science, University of
Ioannina} \centerline{\it P.O.Box 1186, GR-45110 \s Ioannina,
Greece} \centerline{\it e-mail: \{stavros,\s charis\}@cs.uoi.gr}

\vskip 0.5in
\begin{center}
\noindent
\parbox{5.5in}
{{\bf Abstract:} \quad In this paper we examine the classes of
graphs whose $K_n$-complements are trees and quasi-threshold
graphs and derive formulas for their number of spanning trees; for
a subgraph $H$ of $K_n$, the $K_n$-complement of $H$ is the graph
$K_n-H$ which is obtained from $K_n$ by removing the edges of $H$.
Our proofs are based on the complement spanning-tree matrix
theorem, which expresses the number of spanning trees of a graph
as a function of the determinant of a matrix that can be easily
constructed from the adjacency relation of the graph. Our results
generalize previous results and extend the family of graphs of the
form $K_n-H$ admitting formulas for the number of their spanning
trees.

}
\end{center}

\vskip 0.3in 
\section{Introduction}
We consider finite undirected graphs with no loops or multiple
edges. Let $G$ be such a graph on $n$ vertices. A \emph{spanning
tree} of $G$ is an acyclic $(n-1)$-edge subgraph; note that it is
connected and spans $G$. Let $K_n$ denote the complete graph on
$n$ vertices. If $H$ is a subgraph of $K_n$, then $K_n-H$ is
defined to be the graph obtained from $K_n$ by removing the edges
of $H$; the graph $K_n-H$ is called the {\em $K_n$-complement} of
$H$. Note that, if $H$ has $n$ vertices, then $K_n-H$ coincides
with the graph $\skew3\overline{H}$, the {\em complement} of $H$.

\y The problem of calculating the number of spanning trees of a
graph is an important, well-studied problem. Deriving formulas for
different types of graphs can prove to be helpful in identifying
those graphs that contain the maximum number of spanning trees.
Such an investigation has practical consequences related to
network reliability \cite{BrownMPR, Colbourn, MyrCPP, PetBoSuf}.

\y Thus, for both theoretical and practical purposes, we are
interested in deriving formulas for the number of spanning trees
of classes of graphs of the form $K_n - H$. Many cases have
already been examined. For example there exist formulas for the
cases when $H$ is a pairwise disjoint set of edges
\cite{Weinberg}, when it is a star \cite{ONeil}, when it is a
complete graph \cite{Berge}, when it is a path \cite{GilbertMyr},
when it is a cycle \cite{GilbertMyr}, when it is a multi-star
\cite{ChungYan, NikRond, YanMyrChung}, and so on (see Berge
\cite{Berge} for an exposition of the main results).

\y The purpose of this paper is to derive formulas regarding the
number of spanning trees of the graph $G = K_{n} - H$ in the cases
where $H$ is $(i)$ a \emph{tree} on $k$ vertices, $k \leq n$, and
$(ii)$ a \emph{quasi-threshold} graph (or QT-graph for short) on
$p$ vertices, $p \leq n$. A QT-graph is a graph that contains no
induced subgraph isomorphic to $P_{4}$ or $C_{4}$, the path or
cycle on four vertices \cite{Gol1, MaWallisWu, Nik2,
YanChenChang}. Our proofs are based on a classic result known as
the \emph{complement spanning-tree matrix} theorem
\cite{Temperley}, which expresses the number of spanning trees of
a graph $G$ as a function of the determinant of a matrix that can
be easily constructed from the adjacency relation (adjacency
matrix, adjacency lists, etc.) of the graph $G$. Calculating the
determinant of the complement spanning-tree matrix seems to be a
promising approach for computing the number of spanning trees of
families of graphs of the form $K_n - H$, where $H$ posses an
inherent symmetry (see \cite{Berge, ChungYan, GilbertMyr, NikRond,
YanMyrChung, ZaYoGo}). In our cases, since neither trees nor
quasi-threshold graphs possess any symmetry, we focus on their
structural and algorithmic properties. Indeed, both trees and
quasi-threshold graphs possess properties that allow us to
efficiently use the complement spanning-tree matrix theorem; trees
are characterized by simple structures and quasi-threshold graphs
are characterized by a unique tree representation \cite{KanoNik,
Nik2} (see Section 2). We compute the number of spanning trees of
these graphs using standard techniques from linear algebra and
matrix theory on their complement spanning-tree matrices.

\y Various important classes of graphs are trees, including paths,
stars and multi-stars. Moreover, the class of quasi-threshold
graphs contains the classes of perfect graphs known as threshold
graphs and complete split (or, c-split) graphs (see Remark~4.1)
\cite{Gol, HaKe}. Thus, the results of this paper generalize
previous results and extend the family of graphs of the form $K_n
- H$ having formulas regarding the number of spanning trees.

\y The paper is organized as follows. In Section 2 we establish
the notation and related terminology and we present background
results. In particular, we show structural properties for the
class of quasi-threshold graphs and define a unique tree
representation of such graphs. In Sections 3 and 4 we present the
results obtained for the graphs $K_n-T$ and $K_n-Q$, respectively,
where $T$ is a tree and $Q$ is a quasi-threshold graph. Finally,
in Section 5 we conclude the paper and discuss possible future
extensions.

\vskip 0.3in 
\section{Definitions and Background Results}

We consider finite undirected graphs with no loops or multiple
edges. Let $G$ be such a graph; then $V(G)$ and $E(G)$ denote the
set of vertices and of edges of $G$ respectively. The {\it
neighborhood\/}~$N(x)$ of a vertex~$x \in V(G)$ is the set of all
the vertices of $G$ that are adjacent to $x$. The {\it closed
neighborhood\/} of $x$ is defined as $N[x] := \{x\} \cup N(x)$.

\y Let $G$ be a graph on $n$ vertices. The \emph{complement
spanning-tree matrix} $A$ of the graph $G$ is defined as follows:
\begin{equation*}
  A_{i,j}=
  \begin{cases}
    1-\frac{d_i}{n} & \text{if } i=j,\\
    \frac{1}{n}&
    \text{if } i \neq j \text{ and } (i,j) \text{ is not an edge of } G,\\
    0 & \text{otherwise},
  \end{cases}
\end{equation*}

\yy\noindent where $d_i$ is the number of edges incident to vertex
$u_i$ in the complement of $G$; that is, $d_i$ is the degree of
the vertex $u_i$ in $\skew3\overline{G}$. It has been shown
\cite{Temperley} that the number of spanning trees $\tau(G)$ of
$G$ is given by
\begin{equation*}
\tau(G)=n^{n-2} \det(A).
\end{equation*}

\y\noindent In the case where $G=K_n$, we have that $\det(A)=1$;
\emph{Cayley's tree formula} \cite{Harary} states that
$\tau(K_n)=n^{n-2}$.

\y We next provide characterizations and structural properties of
QT-graphs and show that such a graph has a unique tree
representation. The following lemma follows immediately from the
definition of $G[S]$ as the subgraph of $G$ induced by the subset
$S$ of the vertex set $V(G)$.

\vfill\eject

\chl {\bf Lemma~2.1 (\cite{KanoNik, Nik2}).} \s \textit{If $G$ is
a QT-graph, then for every subset $S \subseteq V(G)$, $G[S]$ is
also a QT-graph.} \yy

\y The following theorem provides important properties for the
class of QT-graphs. For convenience, we define
\begin{equation*}
 \text{cent}(G) = \{x \in V(G) \mid N[x] = V(G)\}.
\end{equation*}

\yyy \y
\par \noindent
{\bf Theorem~2.1 (\cite{KanoNik, Nik2}).} \s {\it Let $G$ be an
undirected graph.}
\begin{itemize}
\item[\it (i)]
    {\it $G$ is a QT-graph if and only if every connected
    induced subgraph $G[S], S \subseteq V(G),$ satisfies
    $\text{cent}(G[S]) \neq \emptyset$.}
\item[\it (ii)]
    {\it $G$ is a QT-graph if and only if $G[V(G)-\text{cent}(G)]$
    is a QT-graph.}
\item[\it (iii)]
    {\it Let $G$ be a connected QT-graph.
    If $V(G)-\text{cent}(G)\neq \emptyset$, then $G[V(G)-\text{cent}(G)]$
    contains at least two connected components.}
\end{itemize}

\yy Let $G$ be a connected QT-graph. Then $V_1 := \text{cent}(G)$
is not an empty set by Theorem 2.1. Put $G_1 := G$, and
$G[V(G)-V_1] = G_2 \cup G_3 \cup \cdots \cup G_r$, where each
$G_i$ is a connected component of $G[V(G)-V_1]$ and $r \geq 3$.
Then since each $G_i$ is an induced subgraph of $G$, $G_i$ is also
a QT-graph, and so let $V_i := \text{cent}(G_i) \neq \emptyset$
for $2 \leq i \leq r$. Since each connected component of
$G_i[V(G_i)-\text{cent}(G_i)]$ is also a QT-graph, we can continue
this procedure until we get an empty graph. Then we finally obtain
the following partition of $V(G)$:
\begin{equation*}
    V(G) = V_1 + V_2 + \cdots + V_k, \ss \text{where} \s V_i = \text{cent}(G_i).
\end{equation*}

\chl Moreover we can define a partial order $\preceq$ on  $\{ V_1,
V_2, \ldots, V_k \} $ as follows:
\begin{equation*}
    V_i \preceq V_j \ss \text{if} \ss V_j \subseteq V(G_i).
\end{equation*}

\y \noindent It is easy to see that the above partition of $V(G)$
possesses the following properties.

\yyy \y
\par\noindent
{\bf Theorem~2.2 (\cite{KanoNik, Nik2}).} \s {\it Let $G$ be a
connected QT-graph, and let $V(G) = V_1 + V_2 + \cdots + V_k$ be
the partition defined above; in particular, $V_1 :=
\text{cent}(G)$. Then this partition and the partially ordered set
$(\{V_i\}, \preceq)$ have the following properties:}
\renewcommand {\theenumi} {(P\arabic{enumi})}
\begin{itemize}
\item[\it (P1)]
    \textit{If $V_i \preceq V_j$, then every vertex of $V_i$ and
    every vertex of $V_j$ are joined by an edge of $G$.}
\item[\it (P2)]
    \textit{For every $V_j,
    \text{cent}(G[ \{ \bigcup V_i  \mid V_i \preceq V_j \} ]) =
    V_j$.}
\item[\it (P3)]
    \textit{For every two $V_s$ and $V_t$ such that
    $V_s \preceq V_t$, $G[ \{ \bigcup V_i \mid V_s \preceq V_i \preceq V_t \} ]$
    is a complete graph.
    Moreover, for every maximal element $V_t$ of $(\{ V_i \}, \preceq)$,
    $G[\{ \bigcup V_i  \mid V_1 \preceq V_i \preceq V_t \} ]$ is a maximal
    complete subgraph of $G$.}
\item[\it (P4)]
    \textit{Every edge with both endpoints in $V_i$ is a free
    edge; an edge $(x, y)$ is called free if $N[x] = N[y]$.}
\item[\it (P5)]
    \textit{Every edge with one endpoint in $V_i$ and the other
    endpoint in $V_j$, where $V_i \neq V_j$, is a semi-free
    edge; an edge $(x, y)$ is called semi-free if either
    $N[x] \subset N[y]$ or $N[x] \supset N[y]$.}
\end{itemize}
\yy

\y The results of Theorem 2.2 provide structural properties for
the class of QT-graphs. We shall refer to the structure that meets
the properties of Theorem 2.2 as the \emph{cent-tree} of the graph
$G$ and denote it by $T_c(G)$. The cent-tree is a rooted tree with
root $V_1$; every node $V_i$ of the tree $T_c(G)$ is either a leaf
or has at least two children. Moreover, $V_s \leq V_t$ if and only
if $V_s$ is an ancestor of $V_t$ in $T_c(G)$.

\vskip 0.3in 
\section{Trees}

Let $T$ be a tree on $k$ vertices. In the following construction
we view $T$ as an ordered, rooted tree: one vertex $r \in V(T)$ is
specified as the root and the children of each vertex are given an
ordering (the root is not considered a leaf if it has one child).
We partition the vertex set of the graph $T$, in the following
manner:

\y We set $T_1 := T$ and let leaves$(T_1)$ be the set of leaves of
the tree $T_1$. Then $V_1 := \text{leaves}(T_1)$ is not an empty
set. We delete the leaves of the tree $T_1$ and let $T_2$ be the
resulting tree. We set $V_2 := \text{leaves}(T_2)$ and we continue
this procedure until we get an empty tree. Then, we finally obtain
the following partition of $V(T)$:
\begin{equation*}
    V(T) = V_1 + V_2 + \cdots + V_h,
\end{equation*}
where
\begin{equation*}
    V_i = \text{leaves}(T_i), \ss
    T_{i+1}=T_{i}-\text{leaves}(T_{i}),
    \ss \text{and} \ss T_1 = T.
\end{equation*}

\y \noindent We call this partition the \emph{st-partition} of the
tree $T$.

\y We consider the vertex sets $V_1, V_2, \ldots, V_h$ of the
$st$-partition of a graph $T$ as ordered sets; we here adopt the
left-to-right ordering within $T$. Denote by $V_i^{-1}(u_j)$ the
position of the vertex $u_j$ in the ordered set $V_i$.

\y We label the vertices of $T$ from $1$ to $k$ in the order that
they appear in the ordered sets $V_1, V_2, \ldots, V_h$. More
precisely, if $\ell_i$ and $\ell_j$ denote the labels of the
vertices $u_i$ and $u_j$, respectively, then $\ell_i < \ell_j$ if
and only if either both vertices $u_i$ and $u_j$ belong to the
same vertex set $V_p$ and $V_p^{-1}(u_i) < V_p^{-1}(u_j)$ or
vertices $u_i$ and $u_j$ belong to different vertex sets $V_p$ and
$V_q$, respectively, and $p < q$. This labeling defines a vertex
ordering of $T$; we call it the \emph{st-labeling} of $T$.

\y Let $\ell_1, \ell_2, \ldots, \ell_k$ be the labels taken by the
$st$-labeling of the tree $T$. For every vertex $u_i$ of $T$, we
define the vertex set $\text{ch}(i) \subseteq V(T)$ as follows:
\begin{eqnarray}
    \text{ch}(i) = \{u_j \in V(T) \ss | \ss u_j \in N(u_i) \ss
    \text{and} \ss \ell_i > \ell_j \}. \nonumber
\end{eqnarray}

\y \noindent Hereafter, we shall also use $i$ to denote the vertex
$u_i$ of $T$, $1 \leq i \leq k$. Note that $i \in V(T)$ is a leaf
if and only if $\text{ch}(i) = \emptyset$. Given a rooted tree
$T$, we recursively define the following function $L$ on $V(T)$:

\begin{equation*}
  L(i)=
   \begin{cases}
    a_i & \text{if } i \text{ is a leaf}, \\\\
    a_i-b^2\displaystyle \sum_{j \in \text{ch}(i)} \frac{1}{L(j)} &
    \text{otherwise},
  \end{cases}
\end{equation*}

\yyy\noindent where $a_i = 1 - d_i b$ and $b=1/n$; recall that $n
\geq k$ and $d_i$ is the degree of the vertex $i$ in $T$. We call
$L$ the \emph{st-function} of $T$; hereafter, we use $L_i$ to
denote $L(i)$, $1 \leq i \leq k$.

\y We consider the graph $G=K_n-T$, where $T$ is a tree on $k$
vertices. We first assign to each vertex of the graph $G$ a label
from 1 to $n$ so that the vertices with degree $n-1$ obtain the
smallest labels; that is, we label the vertices with degree $n-1$
from 1 to $n-k$. We label all the other vertices with degree less
than $n-1$ from $n-k+1$ to $n$ according to the $st$-labeling of
$T$. Notice that the vertices with degree less than $n-1$ induce
the graph $\overline{T}$ (note that this is the complement of $T$
in $K_n[T]$, not in $K_n$).

\y Then, we form the complement spanning-tree matrix $A$ of the
graph $G$; it has the following form:

\chl \bs $A=\left[
\begin{array}{cc}
  I_{n-k} &   \\
    & B \\
\end{array}
\right]$,

\chl where the submatrix $B$ concerns those vertices of the graph
$K_n-T$ that have degree less than $n-1$; throughout the paper,
empty entries in matrices or determinants represent zeros. Let

\yyy $V_1 = (u_1, u_2, \ldots, u_\ell)$,

\y $V_2 = (u_{\ell+1}, u_{\ell+2}, \ldots, u_s)$,

\y $V_3 = (u_{s+1}, u_{s+2}, \ldots, u_r)$,

\bs \vdots

\y $V_h = (u_k)$

\chl be the vertex sets of the $st$-partition of $T$; recall that
the vertices $u_1, u_2, \ldots, u_k$ of $K_n-T$ have degrees less
than $n-1$. Thus, $B$ is a $k \times k$ matrix having the
following structure:

\stepcounter{equation}

\chl \bs $B = \left[
\begin{array}{ccccccccccc}
  a_1 &  &   &   &   &   &   &   &   &   &   \\
    & \ddots &   &   &   &   &   &   &   &   &   \\
    &   & a_\ell &   &  &   &   &   &   &   &   \\
    &   &   & a_{\ell+1}  &  &   &   & (b)_{j, i}  &   &   &   \\
    &   &   &   & \ddots  &   &   &   &   &   &   \\
    &   &   &   &   & a_s &  &  &   &   &   \\
    &   &   &   &   &   &  a_{s+1} &  &  &   &   \\
    &   &   & (b)_{i, j}  &   &   &   & \ddots  &   &   &   \\
    &   &   &   &   &   &   &   &  a_r &   &   \\
    &   &   &   &   &   &   &   &   & \ddots  &   \\
    &   &   &   &   &   &   &   &   &   & a_k
\end{array} \right]$, \hspace{\stretch{1}}{(\arabic{equation})\label{tr}}

\chl where, according to the definition of the complement
spanning-tree matrix, $a_i = 1 - d_i b$, and the entries $(b)_{i,
j}$ and $(b)_{j, i}$ of the off-diagonal positions $(i, j)$ and
$(j, i)$ are both $b$ if $j \in \text{ch}(i)$ and 0 otherwise, $1
\leq j \leq i \leq k$. Note that $b=1/n$ and $d_i$ is the degree
of the vertex $i$ in $T$.

\y Starting from the upper left part of the matrix, the first
$\ell$ rows of the matrix correspond to the $\ell$ vertices of the
set $V_1$; the next $s-\ell$ rows correspond to the vertices of
the set $V_2$, and so forth. The last row corresponds to the root
of $T$.

\y From the form of the matrix $A$, we see that $\det(A)=\det(B)$.
Thus, we focus on the computation of the determinant of matrix
$B$.

\y In order to compute the determinant $\det(B)$, we start by
multiplying each column $i$, $1 \leq i \leq \ell$, of the matrix
$B$ by $-b / a_i$ and adding it to the column $j$ if $(b)_{i,j}=b$
($i < j \leq k$). This makes all the strictly upper-diagonal
entries $(b)_{i,j}$, that is, for $i<j \leq \ell$, into zeros. Now
expand in terms of rows $1, 2, \ldots, \ell$, getting

\chl \bs $\det(B) \s = \s \displaystyle \prod_{i=1}^{\ell}{L_i}
\left |
\begin{array}{cccccccc}
    f_{\ell+1}^{\ell} &   &   &   &   &   &   &   \\
    & \ddots  &  &   &   &   &   &   \\
    &   & f_{s}^{\ell} &  &  & (b)_{j, i} &   &   \\
    &   &   &  f_{s+1}^{\ell} &  &  &   &   \\
    &   &   &   & \ddots  &   &   &   \\
    &   & (b)_{i, j}  &   &   &  f_{r}^{\ell} &   &   \\
    &   &   &   &   &   & \ddots  &   \\
    &   &   &   &   &   &   & f_{k}^{\ell}
\end{array}
\right | \ss = \ss \displaystyle \prod_{i=1}^{\ell}{L_i} \s
\det(B')$,

\chl where

\chl \bs $L_i = a_i$, for $1 \leq i \leq \ell$, since the vertices
$1, 2, \ldots, \ell$ are leaves of $T$, and

\chl \bs $f_t^{\ell} = a_t-b^2\displaystyle\sum_{ i \in
\text{ch}(t) \atop 1 \leq i \leq \ell} \frac{1}{L_i}, \sss $for
$\ell+1 \leq t \leq k.$

\yyy \yyy \noindent We observe that the $(k-\ell) \times (k-\ell)$
matrix $B'$ has a structure similar to that of the initial matrix
$B$; see \mbox{Eq.~(1)}. Thus, for the computation of its
determinant $\det(B')$, we follow a similar simplification; that
is, we start by multiplying each column $i$, $1 \leq i \leq
s-\ell$, of the matrix $B'$ by ${-b}/f_i^{\ell}$ and adding it to
the column $j$ if $(b)_{i,j}=b$ ($s < j \leq k$). Then, we obtain

\chl \bs $\det(B) \s = \s \displaystyle \prod_{i=1}^{\ell}{L_i}
\prod_{i=\ell+1}^{s}\!\!\!{L_i} \left |
\begin{array}{ccccc}
    f_{s+1}^{s} &  &  &   &   \\
    & \ddots  &   & (b)_{j, i}  &   \\
    &   &  f_{r}^{s} &   &   \\
    & (b)_{i, j}  &   & \ddots  &   \\
    &   &   &   & f_k^{s}
\end{array}
\right | \ss = \ss \displaystyle \prod_{i=1}^{s}{L_i} \s
\det(B'')$,

\chl where

\chl \bs $L_i = f_i^{\ell}$, for $\ell+1 \leq i \leq s $, and

\chl \bs $f_t^{s} = a_t-b^2\displaystyle\sum_{ i \in \text{ch}(t)
\atop 1 \leq i \leq s } \frac{1}{L_i}, \sss $ for $s+1 \leq t \leq
k.$

\yyy \yyy \noindent The matrix $B''$ also has structure similar to
that of the initial matrix $B$; see \mbox{Eq.~(1)}. It differs
only on the smaller size and on the diagonal values. Thus,
continuing in the same fashion we can finally show that

\chl
\BBs \BBs \Bs \bs $\det(B)=\displaystyle\prod_{i=1}^{k} L_i$,

\chl where $L$ is the $st$-function of $T$ and $k$ is the number
of vertices of $T$.

\y Thus, based on the formula that gives the number $\tau(G)$ of
the spanning trees of the graph $G = K_n - T$ and the fact that
$\det(A)=\det(B)$, we obtain the following result.

\y\chl {\bf Theorem~3.1.} \textit{Let $T$ be a tree on $k$
vertices, $k \leq n$, and let $L$ be the $st$-function on $T$. The
number of spanning trees of the graph $G = K_n - T$ is equal to
\begin{equation*}\label{t1}
  \tau(G)=n^{n-2}\prod_{i=1}^{k} {L_i}.
\end{equation*}
}

\chl {\bf Remark~3.1.} \s We point out that Theorem 3.1 provides a
simple linear-time algorithm for computing the number of spanning
trees of the graph $G = K_n - T$, where $T$ is a tree on $k$
vertices, $k \leq n$; that is, for a graph on $n$ vertices and $m$
edges the algorithm runs in $O(n+m)$ time. Note that the time
complexity is measured according to the uniform cost criterion;
under the uniform cost criterion each instruction requires one
unit of time and each register requires one unit of space. \s
\textbf{$\Box$}

\vskip 0.3in 
\section{Quasi-threshold Graphs}
In this section, we derive a formula for the number of the
spanning trees of the graph $K_n - Q$, where $Q$ is a
quasi-threshold graph.

\y Let $Q$ be a QT-graph on $p$ vertices and let $V_1, V_2
,\ldots,V_k$ be the nodes of its cent-tree $T_c(Q)$ containing
$p_1, p_2, \ldots, p_k$ vertices, respectively. We let $d_i$
denote the degree of an arbitrary vertex of the node $V_i$. Recall
that all the vertices $u \in V(Q)$ of a node $V_i$ have the same
degree. In Figure 1 we show a cent-tree of a QT-graph on 12
vertices. Nodes $V_3$ and $V_{10}$ contain two vertices, while all
the other contain one vertex. The degree of a vertex in node $V_3$
is 4.

\begin{figure}[h] \label{fcent}
  \centering
  \includegraphics{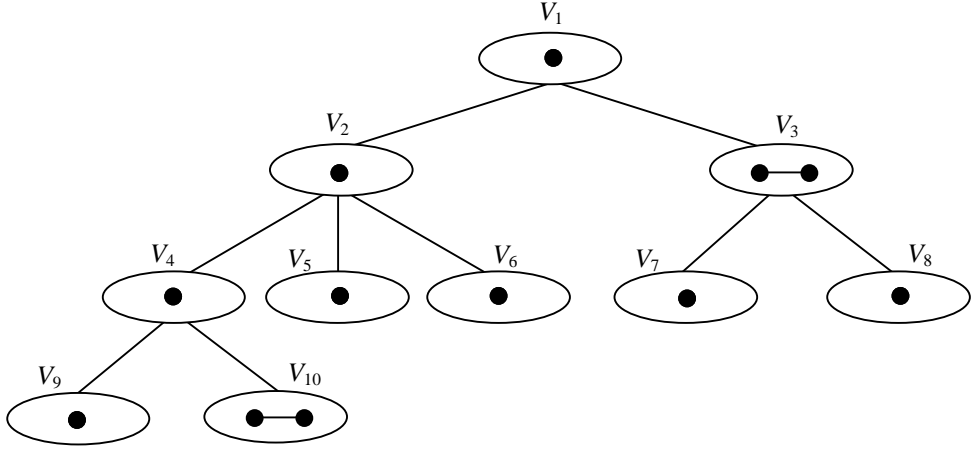}
  \caption{A cent-tree $T_c(Q)$ of a QT-graph on 12 vertices.}
\end{figure}

\yyy\noindent We next form the submatrix $B$ of the complement
spanning-tree matrix $A$ for the graph $K_n-Q$ based on the
structure of the cent-tree $T_c(Q)$, as well as on the
$st$-partition of $T_c(Q)$.

\y Let $V'_1, V'_2, \ldots, V'_h$ be the node sets of the
$st$-partition of $T_c(Q)$. More precisely, the nodes of the
$T_c(Q)$ are partitioned in the following sets:
\begin{eqnarray*}
V'_1 & = & V_1, \ldots, V_{\ell}, \Bs \Bs \Bs \Bs \Bs \Bs \Bs \Bs \Bs \bs \\
V'_2 & = & V_{\ell+1}, \ldots, V_{s},  \Bs \Bs \Bs \Bs \Bs \Bs \Bs \Bs \Bs \bs \\
 & \vdots &  \\
V'_h & = & V_k. \Bs \Bs \Bs \Bs \Bs \Bs \Bs \Bs \Bs \bs
\end{eqnarray*}
\noindent Then, we label the vertices of the graph $Q$ from
$n-p+1$ to $n$ as follows: First, we label the vertices in $V_{1}$
from $(n-p)+1$ to $(n-p)+p_1$; next, we label the vertices in
$V_{2}$ from $(n-p)+p_1+1$ to $(n-p)+p_1+p_2$; finally, we label
the vertices in $V_{k}$.

\y Thus, based on the above labeling of the vertices of the
QT-graph $Q$, we can easily construct the matrix $B$ of the graph
$K_n-Q$; it is a $p \times p$ matrix and has the following form:

\stepcounter{equation}

\chl \bs $B=\left [\begin{array}{ccccccccccc}
   M_1 &   &   &   &   &   &   &   &   &   &   \\
    & \ddots  &   &   &   &   &   &   &   &   &   \\
    &   & M_\ell  &   &   &   &   &   &   &   &   \\
    &   &   & M_{\ell+1}  &   &   &   & [b]_{j, i}  &   &    &   \\
    &   &   &   & \ddots  &   &   &   &   &    &   \\
    &   &   &   &   & M_s  &   &   &   &   &   \\
    &   &   &   &   &   &  M_{s+1} &   &   &   &   \\
    &   &   &[b]_{i, j}   &   &   &   &  \ddots &   &   &   \\
    &   &   &   &   &   &   &   & M_{r}  &   &   \\
    &   &   &   &   &   &   &   &   & \ddots  &   \\
    &   &   &   &   &   &   &   &   &   & M_{k}
\end{array}
\right]$, \hspace{\stretch{1}}{(\arabic{equation})\label{QTb}}

\chl where $M_i$ is a $p_i \times p_i$ submatrix of the form

\chl \bs $M_i= \left[
\begin{array}{cccc}
  a_{i} & b & \cdots & b \\
  b & a_{i} & \cdots & b \\
  \vdots & \vdots & \ddots & \vdots \\
  b & b & \cdots & a_{i}
\end{array} \right]$,

\chl and the entries $[b]_{i, j}$ and $[b]_{j, i}$ of the
off-diagonal positions $(i, j)$ and $(j, i)$, respectively, of
matrix $B$ correspond to $p_i \times p_j$ and $p_j \times p_i$
submatrices with all their elements $b'$s if node $V_j$ is a
descendant of node $V_i$ in $T_c(Q)$ and zeros otherwise, $1 \leq
j \leq i \leq k$. Recall that $a_{i} = 1-d_i b$, where $d_i$ is
the degree of an arbitrary vertex in node $V_i$ of $T_c(Q)$, and
$b = 1/n$.

\y In order to compute the determinant of the matrix $B$ we first
simplify the determinants of the matrices $M_i$, $1 \leq i \leq
k$. We multiply the last row of the matrix $M_i$ by $-1$ and add
it to the first $p_i-1$ rows of the matrix $M_i$, $1 \leq i \leq
k$. Then we add the first $p_i-1$ columns of the matrix $M_i$ to
the last column of the matrix $M_i$, $1 \leq i \leq k$, and we
obtain

\chl \bs $\det(M_i)= \left |\begin{array}{cccc}
  a_{i}-b &  &  &  \\
   & a_{i}-b &  &  \\
   &  & \ddots &  \\
  b & b &  & a_{i}-b+p_ib
\end{array}
 \right | = (a_{i}-b)^{p_i-1}(a_{i}-(1-p_i)b)$.

\yy \chl It now suffices to substitute the above value in the
determinant of matrix $B$. We point out that after simplifying the
determinant of matrices $M_i$ only the diagonal and the last row
of each matrix $M_i$ have nonzero entries; the diagonal has
nonzero entries since $d_i < n-1$. Thus, we have

\begin{equation}\label{dd}
    \det(B)=
    \displaystyle \prod_{i=1}^{k}p_i({a_{i}-b})^{p_i-1} \det(D),
\end{equation}

\stepcounter{equation}

\noindent where
\chl \bs $D=\left[
\begin{array}{ccccccccccc}
   \sigma_{1} &   &   &   &   &   &   &   &   &   &   \\
    & \ddots  &   &   &   &   &   &   &   &   &   \\
    &   &  \sigma_{\ell} &   &   &   &   &   &   &   &   \\
    &   &   & \sigma_{\ell+1}  &   &   &   & (b)_{j, i}  &   &    &   \\
    &   &   &   & \ddots  &   &   &   &   &    &   \\
    &   &   &   &   & \sigma_{s} &   &   &   &   &   \\
    &   &   &   &   &   & \sigma_{s+1}  &   &   &   &   \\
    &   &   & (b)_{i, j}  &   &   &   & \ddots  &   &   &   \\
    &   &   &   &   &   &   &   & \sigma_{r} &   &   \\
    &   &   &   &   &   &   &   &   &  \ddots &   \\
    &   &   &   &   &   &   &   &   &   &  \sigma_{k}
\end{array} \right]
$ \hspace{\stretch{1}}{(\arabic{equation})\label{DD1}}

\yy \chl is a $k \times k$ matrix with diagonal elements $\sigma_i
= \frac{a_{i}-(1-p_{i})b}{p_{i}}$, $1 \leq i \leq k$, and the
entry $(b)_{i, j}$ of the off-diagonal position $(i, j)$ is $b$ if
node $V_j$ is a descendant of node $V_i$ in $T_c(Q)$ and $0$
otherwise, $1 \leq j \leq i \leq k$.

\y We observe that if we set $p_i = 1$ in matrix $D$, $1 \leq i
\leq k$, then $D$ is equal to the submatrix $B$ of the graph
$K_n-Q$, where $Q$ is a graph of a special type; it is a QT-graph
on $k$ vertices possessing the property that each node of its
cent-tree $T_c(Q)$ contains a single vertex; see \mbox{Figure 2}.

\begin{figure}[h] \label{fff}
  \centering
  \includegraphics{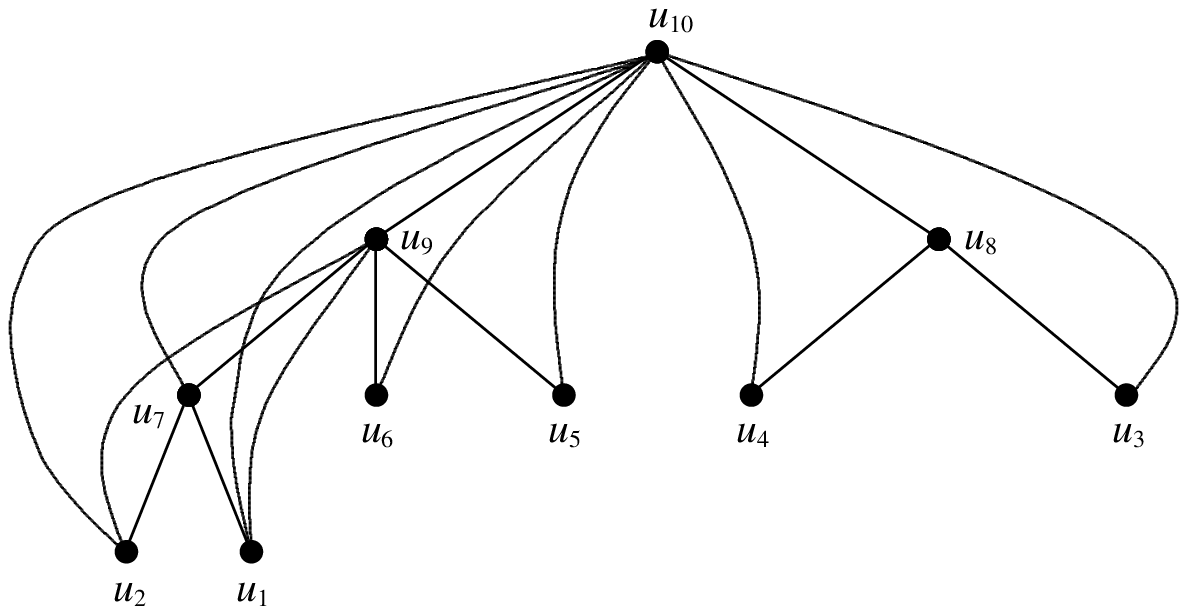}
  \captionstyle{center}
  \onelinecaptionsfalse
  \caption{A QT-graph $Q$ on 10 vertices.
  Every node $V_i$ of the cent-tree $T_c(Q)$ \protect\\
  contains exactly one vertex.}
\end{figure}

\y It is easy to see that, if we form the submatrix $B$ of the
complement spanning-tree matrix $A$ of $K_n-Q$, where $Q$ is the
QT-graph of \mbox{Figure 2}, using an appropriate vertex labeling,
that is, $\ell_2 = n-9, \s \ell_1 = n-8, \s \ldots, \s \ell_{10} =
n$, then we obtain $D = B$. The idea now is to transform the $k
\times k$ matrix $D$ into a form similar to that of the $k \times
k$ matrix $B$ of a tree $T$ on $k$ vertices; see Eq. (1) in
Section 3. We proceed as follows:

\chl We first apply the following operations to each row $i=1, 2,
\ldots, k$ of the matrix $D$:

\begin{itemize}
    \item[$\bullet$]
        We find the minimum index $j$ such that $i < j \leq k$ and
        $D_{i,j} \neq 0$, and then
    \item[$\bullet$]
        we multiply the $j$th column by $-1$ and add it to the $\ell$th column,
        if $D_{i,\ell}=D_{i,j}$ and $j+1 \leq \ell \leq k$.
\end{itemize}

\noindent Next, we apply similar operations to each column $j=1,
2, \ldots, k$ of the matrix $D$:

\begin{itemize}
    \item[$\bullet$]
        We find the minimum index $i$ such that $1 \leq j < i$ and
        $D_{i,j} \neq 0$, and then
    \item[$\bullet$]
        we multiply the $i$th row by $-1$ and add it to the $\ell$th row,
        if $D_{\ell,j}=D_{i,j}$ and $i+1 \leq \ell \leq k$.
\end{itemize}

\y\noindent Thus, we obtain

\chl \bs $\det(D)=\left | \begin{array}{ccccccccccc}
    a'_{1} &   &   &   &   &   &   &   &   &   &  \\
    & \ddots  &   &   &   &   &   &   &   &   &  \\
    &   & a'_{\ell}  &   &   &   & (b'_j)_{j, i}  &   &   &   &   \\
    &   &   & a'_{\ell+1}  &   &   &   &   &   &   &   \\
    &   &   &   & \ddots  &   &   &   &   &   &   \\
    &   &   &   &   & a'_{s}  &   &   &   &   &   \\
    &   & (b'_j)_{i, j}  &   &   &   & a'_{s+1}  &   &   &   &   \\
    &   &   &   &   &   &   & \ddots  &   &   &   \\
    &   &   &   &   &   &   &   & a'_{r}  &   &   \\
    &   &   &   &   &   &   &   &   & \ddots  &   \\
    &   &   &   &   &   &   &   &   &   &  a'_{k}
\end{array}
\right|$,

\chl where
\begin{eqnarray}{\label{aa'}}
    a'_{i}=
    \begin{cases}
        \sigma_i & \text{if} \s V_i \s \text{is a leaf of} \s T_c(Q), \\\\
        \sigma_i+\displaystyle\sum_{j \in \text{ch}(i) \atop \ell+1 \leq j \leq k} (\sigma_j-2b) &
        \text{otherwise},
    \end{cases}
\end{eqnarray}

\chl and
\begin{eqnarray}{\label{bb'}}
    b'_i=
    \begin{cases}
        b & \text{if } V_i \text{ is a leaf of }T_c(Q), \\\\
        b-\sigma_i & \text{otherwise}.
    \end{cases}
\end{eqnarray}

\chl Note that the entry $(b'_j)_{i, j}$ in the off-diagonal
position $(i, j)$ is $b'_j$ if node $V_j$ is a descendant of node
$V_i$ in $T_c(Q)$ and $0$ otherwise, $1 \leq j \leq i \leq k$.
Recall that $\sigma_i = \frac{a_{i}-(1-p_{i})b}{p_{i}}$; in the
case where each node of the cent-tree $T_c(Q)$ contains a single
vertex, we have $\sigma_i = a_i$ (in this case $p_i = 1$, for
every $i = 1, 2, \ldots, k$).

\y It is easy to see that the structure of the resulting $k \times
k$ matrix $D$ is similar to that of the $k \times k$ matrix $B$ of
a tree; see \mbox{Eq.~(1)} in Section 3. Thus, for the computation
of the determinant $\det(D)$, we can use similar techniques.

\y We next define the following function $\phi$ on the nodes on
the cent-tree of a QT-graph $Q$:

\begin{eqnarray*}
    \phi(i)=
    \begin{cases}
        a'_i &
        \text{if } i \in V_i \text{ and } V_i \text{ is a leaf of }T_c(Q), \\\\
        a'_i- \displaystyle \sum_{j \in \text{ch}(i)} {\frac{(b'_{j}) ^{2}}{\phi(j)}} &
        \text{otherwise},
    \end{cases}
\end{eqnarray*}

\noindent where $a'_i$ and $b'_i$ are defined in Eq.~(\ref{aa'})
and Eq.~(\ref{bb'}), respectively. We call the function $\phi$ the
\emph{cent-function} of the graph $Q$ or, equivalently, the
\emph{cent-function} of the cent-tree $T_c(Q)$; hereafter, we use
$\phi_i$ to denote $\phi(i)$, $1 \leq i \leq k$.

\y Following the same elimination scheme as that for the
computation of the determinant of the matrix $B$ in Section 3, we
obtain
\begin{eqnarray}{\label{ddphi}}
    \det(D) = \displaystyle \prod_{i=1}^{k} \phi_i.
\end{eqnarray}

\y\noindent Thus, the results of this section are summarized in
the following theorem.

\y\chl {\bf Theorem~4.1.} \textit{Let $Q$ be a quasi-threshold
graph on $p$ vertices and let $V_1, V_2, \ldots, V_k$ be the nodes
of the cent-tree of $Q$. Let $\phi$ be the cent-function of the
graph $Q$. Then, the number of spanning trees of the graph $G =
K_n - Q$ is equal to
\begin{equation*}\label{t2}
  \tau(G)=n^{n+k-p-2}\prod_{i=1}^{k} {p_i(n-d_i-1)^{p_i-1}\phi_i},
\end{equation*}
where $p_i$ is the number of vertices of the node $V_i$ and $d_i$
is the degree of an arbitrary vertex in node $V_i$, $1 \leq i \leq
k$.}

\chl {\it Proof}. As mentioned in Section 3, the complement
spanning-tree matrix $A$ of a graph $K_n-Q$ can be represented by

\chl \BBs \BBs \Bs $A=\left[
\begin{array}{cc}
    I_{n-p} & \s  \\
    \s & B \\
\end{array}
\right]$,

\chl where the submatrix $B$ concerns those vertices of the graph
$K_n-Q$ that have degree less than $n-1$; these vertices induce
the graph $\skew3\overline{Q}$. Since $a_i=1-d_ib$ and $b=1/n$,
from \mbox{Eq.~(3)} we have

\begin{equation*}
   \det(B) = n^{k-p}\displaystyle \prod_{i=1}^{k}p_i(n-d_i-1)^{p_i-1}
   \det(D).
\end{equation*}

\chl From the above equality and \mbox{Eq.~(7)}, we obtain

\begin{equation*}
  \det(B) = n^{k-p}\prod_{i=1}^{k} {p_i(n-d_i-1)^{p_i-1}\phi_i}.
\end{equation*}

\chl The number of spanning trees $\tau(G)$ of the graph $G$ is
equal to $n^{n-2}\det(A)$. Thus, since $\det(A)=\det(B)$, the
theorem follows. \qed

\yyy Theorem 4.1 coupled with Theorem 3.1 implies a simple
linear-time algorithm for computing the number of spanning trees
of the graph $G = K_n - Q$, where $Q$ is a quasi-threshold graph
on $p$ vertices, $p \leq n$ (see also Remark 3.1).

\yy\chl {\bf Remark~4.1.} As mentioned in the introduction, the
class of quasi-threshold graphs contains the class of c-split
graphs (complete split graphs); recall that a graph is defined to
be a c-split graph if there is a partition of its vertex set into
a stable set $S$ and a complete set $K$ and every vertex in $S$ is
adjacent to all the vertices in $K$ \cite{Gol}.

\y Thus, the cent-tree of a c-split graph $H$ consists of $|S|+1$
nodes $V_1, V_2, \ldots, V_{|S|+1}$ such that $V_1 = K$ and the
nodes $V_2, V_3, \ldots, V_{|S|+1}$ are children of the root
$V_1$; each child contains exactly one vertex $u \in S$.

\y Let $H$ be a c-split graph on $p$ vertices and let $V(H) = K +
S$ be the partition of its vertex set. Then, by Theorem 4.1, we
obtain that the number of spanning trees of the graph $G=K_n-H$ is
given by the following closed formula: \yyy

\BBs \Bs \bs  $\tau(G)=n^{n-p-1}(n-|K|)^{|S|-1}(n-p)^{|K|}$,

\yy\noindent where $p=|K|+|S|$ and $p \leq n$. \s \textbf{$\Box$}

\vskip 0.3in  
\section{Concluding Remarks}

It is well known that the classes of quasi-threshold and threshold
graphs are perfect graphs. Thus, it is reasonable to ask whether
the complement spanning-tree matrix theorem can be efficiently
used for deriving formulas, regarding the number of spanning
trees, for other classes of perfect graphs \cite{Gol}.

\y It has been shown that the classes of perfect graphs, namely
complement reducible graphs, or so-called cographs, and
permutation graphs, have nice structural and algorithmic
properties: a cograph admits a unique tree representation, up to
isomorphism, called a cotree \cite{Lerchs} (note that the class of
cographs contain the classes of quasi-threshold and threshold
graphs), while a permutation graph $G[\pi]$ can be transformed
into a directed acyclic graph and, then, into a rooted tree by
exploiting the inversion relation on the elements of the
permutation $\pi$ \cite{Nik1}.

\y Based on these properties, one can work towards the
investigation whether the classes of cographs and permutation
graphs belong to the family of graphs that admit formulas for the
number of their spanning trees.

\end{document}